\documentclass[lettersize,journal]{IEEEtran}
\usepackage{amsmath,amsfonts}
\usepackage{algorithmic}
\usepackage{array}
\usepackage[caption=false,font=normalsize,labelfont=sf,textfont=sf]{subfig}
\usepackage{textcomp}
\usepackage{stfloats}
\usepackage{url}
\usepackage{verbatim}
\usepackage{graphicx}
\usepackage{tikz}
\usepackage{tabularx}
\usepackage{booktabs}
\usepackage{amsmath}
\usepackage{multirow}
\usepackage{rotating}
\usepackage{pdflscape}
\usepackage{hyperref}
\usepackage{graphicx}
\usepackage{threeparttable}
\usetikzlibrary{shapes.geometric, arrows}
\usepackage{scalerel}
\usepackage{tikz}
\usetikzlibrary{svg.path}

\definecolor{orcidlogocol}{HTML}{A6CE39}
\tikzset{
  orcidlogo/.pic={
    \fill[orcidlogocol] svg{M256,128c0,70.7-57.3,128-128,128C57.3,256,0,198.7,0,128C0,57.3,57.3,0,128,0C198.7,0,256,57.3,256,128z};
    \fill[white] svg{M86.3,186.2H70.9V79.1h15.4v48.4V186.2z}
                 svg{M108.9,79.1h41.6c39.6,0,57,28.3,57,53.6c0,27.5-21.5,53.6-56.8,53.6h-41.8V79.1z M124.3,172.4h24.5c34.9,0,42.9-26.5,42.9-39.7c0-21.5-13.7-39.7-43.7-39.7h-23.7V172.4z}
                 svg{M88.7,56.8c0,5.5-4.5,10.1-10.1,10.1c-5.6,0-10.1-4.6-10.1-10.1c0-5.6,4.5-10.1,10.1-10.1C84.2,46.7,88.7,51.3,88.7,56.8z};
  }
}

\newcommand\orcidicon[1]{\href{https://orcid.org/#1}{\mbox{\scalerel*{
\begin{tikzpicture}[yscale=-1,transform shape]
\pic{orcidlogo};
\end{tikzpicture}
}{|}}}}

\usepackage{hyperref} %<--- Load after everything else

\IEEEoverridecommandlockouts
\begin{document}

\title{HPCNeuroNet: Advancing Neuromorphic Audio Signal Processing with Transformer-Enhanced Spiking Neural Networks}
\author{ Murat Isik  \orcidicon{0000-0002-0907-7253}, \IEEEmembership{Member, IEEE},
        Hiruna Vishwamith \orcidicon{0009-0004-3037-8554} \IEEEmembership{Member, IEEE}, Kayode Inadagbo \orcidicon{0009-0009-9512-3321} \IEEEmembership{Member, IEEE},
        and I. Can Dikmen, \orcidicon{0000-0002-7747-7777} \IEEEmembership{Senior Member, IEEE}
        
\thanks{M. Isik is with Stanford University, Stanford, CA, USA (e-mail: mrtisik@stanford.edu).}
\thanks{H. Vishwamith is with University Of Moratuwa, Moratuwa, Sri Lanka (e-mail: vishwamithpgh.20@uom.lk).}
\thanks{K. Inadagbo is with Prairie View A\&M University, Texas, USA (e-mail: kayodeinadagbo@gmail.com).}
\thanks{I. C. Dikmen is with Temsa Research\&Development Center, Adana, Turkey (e-mail: can.dikmen@temsa.com).}}

\markboth{Journal of IEEE Transactions on Signal Processing,~Vol.~XX, No.~XX, Month~Year}%
{Isik \MakeLowercase{\textit{et al.}}: HPCNeuroNet: Advancing Neuromorphic Audio Signal Processing}

\maketitle

\begin{abstract}
This paper presents a novel approach to neuromorphic audio processing by integrating the strengths of Spiking Neural Networks (SNNs), Transformers, and high-performance computing (HPC) into the HPCNeuroNet architecture. Utilizing the Intel N-DNS dataset, we demonstrate the system's capability to process diverse human vocal recordings across multiple languages and noise backgrounds. The core of our approach lies in the fusion of the temporal dynamics of SNNs with the attention mechanisms of Transformers, enabling the model to capture intricate audio patterns and relationships. Our architecture, HPCNeuroNet, employs the Short-Time Fourier Transform (STFT) for time-frequency representation, Transformer embeddings for dense vector generation, and SNN encoding/decoding mechanisms for spike train conversions. The system's performance is further enhanced by leveraging the computational capabilities of NVIDIA's GeForce RTX 3060 GPU and Intel's Core i9 12900H CPU. Additionally, we introduce a hardware implementation on the Xilinx VU37P HBM FPGA platform, optimizing for energy efficiency and real-time processing. The proposed accelerator achieves a throughput of 71.11 Giga-Operations Per Second (GOP/s) with a 3.55 W on-chip power consumption at 100 MHz. The comparison results with off-the-shelf devices and recent state-of-the-art implementations illustrate that the proposed accelerator has obvious advantages in terms of energy efficiency and design flexibility. Through design-space exploration, we provide insights into optimizing core capacities for audio tasks. Our findings underscore the transformative potential of integrating SNNs, Transformers, and HPC for neuromorphic audio processing, setting a new benchmark for future research and applications.
\end{abstract}


\section{Introduction}

\IEEEPARstart{H}{igh-Performance} Computing (HPC) has become a cornerstone in the evolution of neuromorphic computing, particularly in the realm of audio processing. The recent surge in neuromorphic hardware architectures, epitomized by platforms like HPCNeuroNet, underscores the pivotal role of HPC in pushing the boundaries of computational capabilities. Neuromorphic systems, inspired by the intricate workings of the human brain, are uniquely positioned to transform the landscape of audio processing. Their event-driven computational approach mirrors biological neural activities, making them particularly adept at handling audio data. However, the journey towards harnessing the full potential of neuromorphic computing is not without its challenges. One of the most pressing issues is the reliance on external software for training neuromorphic systems. This method is not only time-intensive but also consumes significant energy, making it a less-than-ideal solution for real-time applications. The need for more efficient, HPC-driven solutions is evident.

Furthermore, the escalating demand for processing vast datasets, especially in high-resolution audio processing, calls for systems that combine speed, adaptability, and energy efficiency. Field-Programmable Gate Arrays (FPGAs) have emerged as a beacon in this context. Their inherent adaptability, parallelism, and configurability make them ideal for handling the throughput rates required by modern applications. Over the past decade, FPGAs have seen widespread adoption due to their versatility across a myriad of applications. Their inherent flexibility, parallelism, and configurability make them well-suited to meet the rigorous throughput requirements of modern audio applications. Over the past decade, FPGAs have witnessed widespread adoption, a testament to their adaptability across a spectrum of applications. In the current technological landscape, FPGA architectures, such as the one underpinning the HPCNeuroNet system, are increasingly recognized as a viable alternative to traditional FPGAs. These architectures seamlessly integrate the robust performance of FPGAs with the versatility of processors, offering an optimal solution for demanding audio processing applications. While there have been strides in leveraging FPGAs for various applications, the challenge of power consumption remains. This highlights the indispensable role of HPC in achieving performance enhancements while maintaining energy frugality.

\subsection{Contributions of Our Research}

This manuscript delves into the synergy between neuromorphic audio processing and high-performance computing, emphasizing the role of heterogeneous devices. Our primary contributions are:

\begin{itemize}

\item Introduction of the HPCNeuroNet architecture, a novel approach that integrates HPC, SNN dynamics, and Transformer mechanisms for audio processing.

\item Exploration of FPGA architectures in neuromorphic audio processing, detailing the implementation of HPCNeuroNet on FPGA and its benefits.

\item Insights into the combined potential of SNN and Transformer architectures for audio processing, with a focus on real-world applications like audio denoising.

\item Analysis of neuromorphic hardware integration into devices like GPUs, CPUs, and FPGAs, highlighting their respective advantages and performance metrics.

\end{itemize}

\subsection{Organization}
This manuscript is structured as follows: \textbf{Section 2} provides background on neuromorphic computing, its evolution, and its intersection with high-performance computing. \textbf{Section 3} delves into the specifics of neuromorphic audio processing, discussing the methodologies, challenges, and advancements in the field. \textbf{Section 4} presents the methodological approach, detailing the datasets used, the expressivity of Spiking Neural Networks, and the hardware architectures employed. \textbf{Section 5} showcases the results, emphasizing the performance gains achieved through HPC and the scalability of the proposed methods. \textbf{Section 6} offers a discussion on the results, analyzing them in the broader context of HPC and neuromorphic audio processing. \textbf{Section 7} concludes the manuscript, summarizing the main points, and emphasizing the significant contributions and findings of this study. \textbf{Section 8}  discusses potential future directions in neuromorphic audio processing with HPC.

\section{Neuromorphic Computing, High-Performance Computing, and Audio Processing}

\subsection{Evolution of Neuromorphic Computing and Intersection with HPC}

Neuromorphic computing, rooted in brain-inspired principles, has undergone significant evolution. While its core remains centered on emulating the human brain's functionalities, the methodologies have diversified. Techniques such as in-memory computing, continuous learning in hardware, and fine-grained parallelism have been explored. The objectives of neuromorphic research can be categorized into two main streams: understanding biological nervous systems and leveraging brain-inspired principles for efficient computing. The intersection of these objectives is evident in the incorporation of biologically credible mechanisms, potentially surpassing traditional machine learning paradigms. At the algorithmic forefront, neuromorphic systems predominantly employ spiking SNNs or models trained using biologically plausible learning rules. Recent advancements, such as e-prop and decolle, have addressed challenges of spatial and temporal locality in error signals, enabling large-scale SNNs like SpikeGPT \cite{zhu2023spikegpt, zhang2022spiking}. Concurrently, the hardware community has been focusing on neuromorphic devices that leverage material physics to emulate biological functionalities, ranging from memristive to photonic devices. The integration of neuromorphic architectures with HPC has been pivotal in optimizing resource utilization. The event-driven nature of neuromorphic systems, characterized by neural 'spikes', offers energy efficiency, making them ideal for HPC. However, programming challenges persist, especially for realizing a broad class of probabilistic algorithms in SNNs.

\subsection{Role of HPC in Current Audio Processing Techniques}

Neuromorphic computers, with their architecture mimicking the human brain, hold transformative potential for healthcare, especially in point-of-care devices. However, their adoption has been constrained by training challenges, primarily the dependence on external software, which is both cumbersome and energy-inefficient. The integration of neuromorphic systems with HPC can address these limitations, offering enhanced computational power and energy efficiency. In the realm of audio processing, neuromorphic systems combined with HPC can revolutionize techniques, as demonstrated in \cite{cerezuela2016sound}, where an SNN-based sound recognition system achieved over 91\% accuracy even in noisy environments. Similarly, the work in \cite{gao2019real} proposed a low-latency voice recognition system suitable for IoT applications, highlighting the potential of integrating neuromorphic systems, HPC, and audio processing.

\subsection{Network Architectures in HPC Contexts}

The architecture of neural networks plays a significant role in determining their efficiency and applicability in HPC contexts. For instance, edge detection, a critical machine learning task, requires extensive neurons and software simulations, leading to scalability issues and increased computational time. The research in \cite{glackin2009emulating} proposed a scalable FPGA implementation strategy to address these challenges. Researchers \cite{caron2011fpga} employed the Oscillatory Dynamic Link Matcher (ODLM) methods to identify patterns, underscoring the significance of network structures in high-performance computing.

\section{Neuromorphic Audio}
The potential of neuromorphic hardware in audio processing is underscored by its promise in energy efficiency and computational performance. This is where HPC plays a pivotal role. By leveraging the parallel processing capabilities of HPC, neuromorphic audio processing can handle the computational demands of real-time audio recognition, filtering, and classification. A salient advantage of integrating HPC with neuromorphic systems lies in the reduction of model-loading overhead, significantly enhancing processing speeds. Furthermore, the introduction of probabilistic neuromorphic programming, combined with hyper-dimensional computing and vector symbolic architectures like Spatial Semantic Pointers, further augments the capabilities of neuromorphic audio processing.

Neuromorphic architectures, with their inherent parallelism, interconnectedness, and adaptability, are adept at executing intricate operations in real-time \cite{schaik2015silicon}. They strike a balance between quality, efficiency, and resource utilization. These systems employ signals made up of brief temporal pulses, known as spikes or events, with data encoded in various ways, from spike polarity and frequency to time-between-spikes \cite{indiveri2006vlsi, thorpe2010suggestions}. The Address-Event Representation (AER), a notable advancement introduced by Mead lab in 1991 \cite{boahen1998communicating}, facilitates the interconnection of silicon neurons across chips representing different neural layers.

For audio recognition, neuromorphic tools emulate the biological cochlea's function, breaking down an audio signal into various frequency bands of spiking data \cite{schaik2015silicon}. This mirrors the cochlea's conversion of pressure signals into neural signals directed to the brain. The inaugural silicon cochlea, conceptualized by Lyon and Mead \cite{lyon1988analog}, emulated the basilar membrane using a sequence of filter segments. Numerous VLSI adaptations of this model exist \cite{wen2009silicon, hamilton2008active, liu2010event}, with digital cochlea models employing traditional Digital Signal Processing methods \cite{leong2003fpga, dundur2008digital, thakur2014fpga}.

\begin{figure}[h]
    \centering
    \includegraphics[width = 0.5\textwidth]{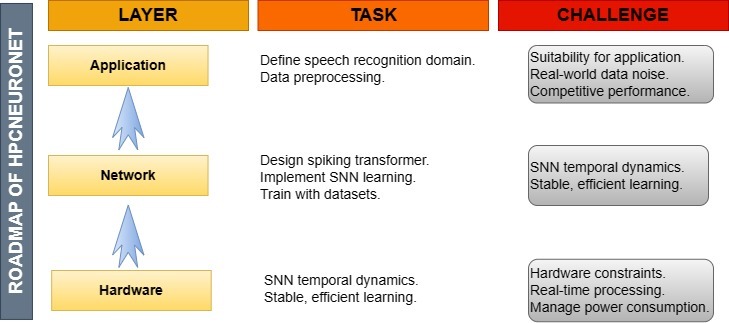}
        \caption{HPCNeuroNet framework depicted across three dimensions: layers, tasks, and challenges of the design.}
        \label{roadmap}
\end{figure}

The Neuromorphic Audio System (NAS) stands out due to its unique approach \cite{DominguezMorales2011, JimenezFernandez2010, GomezRodriguez2005, CerezuelaEscudero2013}. It processes data directly encoded as spikes using Pulse Frequency Modulation (PFM), implementing Spike Signal Processing methods and leveraging AER interfaces. The system ingests digital audio streams, symbolizing monaural sound signals. A Synthetic Spike Generator transforms this input into a spike sequence, which is then divided into distinct frequency bands using a cascading band-pass filter bank. These are amalgamated into an AER output bus, encoding each spike as per AER, and forwarding this data to classification mechanisms. The computational demands of neuromorphic audio, given its intricate operations and real-time requirements, necessitate the power and parallelism of HPC, ensuring optimal performance and efficiency. A roadmap diagram is shown in ~\autoref{roadmap}.

\section{Method}

\subsection{Dataset}

The Intel N-DNS collection, adapted from the Microsoft DNS Challenge, encompasses diverse human vocal recordings in multiple languages and noise backgrounds \cite{xing2020new}. It offers a generation script producing 30-second segments of clean, noise, and noisy audio for training and validation. Leveraging HPC, the collection ensures rapid data processing, especially for large datasets. Contributors can customize generation parameters, choose specific languages and noises, or add other datasets. The collection provides 500 hours of audio with SNRs between 20 dB and -5 dB, recorded at 16 kHz and 16-bit resolution. The validation set maintains standard parameters for consistency. Evaluation data will be released after model submissions, ensuring unbiased assessment. The challenge encourages continuous engagement, positioning it as a dynamic platform for neuromorphic research. Additionally, the challenge includes HPC-optimized dataloaders for efficient audio retrieval, with optional metadata detailing audio origins, noise types, and blend levels.

\subsection{Expressivity of Spiking Neural Networks}

SNNs have emerged as a promising avenue in the realm of neural computation, primarily due to their ability to mimic the temporal intricacies of biological neural systems. Recent studies have unveiled fascinating similarities and contrasts between SNNs and their counterparts, the ReLU-activated artificial neural networks (ReLU-ANNs). This section offers insights into the functional realization capabilities of these networks and underscores their expressivity nuances.

\subsubsection{Function Dissection}
\leavevmode\\
Consider a function, denoted as \( f(x) \), which is articulated as:
\begin{equation}
f(x) = 
\begin{cases} 
x & \text{for } x \leq -\theta, \\
\frac{x-\theta}{2} & \text{for } -\theta < x < \theta, \\
0 & \text{for } x \geq \theta. 
\end{cases}
\end{equation}
This function can also be rephrased using the ReLU activation:
\begin{equation}
f(x) = -\frac{1}{2} \sigma(-x - \theta) - \frac{1}{2} \sigma(-x + \theta),
\end{equation}
where \( \sigma \) epitomizes the ReLU function.

The function \( f(x) \) can be adeptly realized through an SNN, characterized by a set of weights, delays, and thresholds. This network configuration encompasses two input neurons and a singular output neuron. Depending on the magnitude and sign of \( x \), the firing dynamics of the output neuron are influenced by the input spikes. This realization holds true across the entire domain of \( x \).

\subsubsection{Complexity Insights}
\leavevmode\\
In the realm of neural network architectures, the function \( f(x) \) is represented in a two-layered ReLU-ANN. This architecture comprises two hidden units and a singular output unit, with each connection defined by specific weight and bias parameters. The simplicity of the ReLU-ANN structure is evident, especially when considering the tri-linear nature of \( f(x) \). Given that a ReLU unit can segment the input space into a maximum of two linear zones, the necessity for at least two hidden units becomes clear. This emphasizes the efficiency of the ReLU-ANN in capturing \( f(x) \). Transitioning to SNNs, while their inherent delays weren't the focal point of our discussion, they play a crucial role in adjusting spike dynamics, which in turn influences the neural output. A key insight from our exploration is the capability of SNNs to emulate the dynamics of a conventional ReLU network. The function \( f(x) \) exemplifies this, suggesting scenarios where SNNs could potentially surpass ReLU-ANNs in representational efficiency. SNNs, characterized by their transmission of information through spikes or action potentials, are aptly suited for neuromorphic hardware implementations. Recent studies have illuminated the versatility of SNNs, showcasing their ability to emulate Turing machines, arbitrary threshold circuits, and even sigmoidal neurons. Their prowess extends to approximating continuous functions with remarkable precision, underscoring their applicability across diverse domains. In the context of neuromorphic audio processing, restructuring models are pivotal. Strategic deployment, such as utilizing fewer cores, can enhance both energy efficiency and computational throughput. Furthermore, the synergy between neuromorphic hardware architectures and high-performance computing platforms promises substantial benefits in audio processing. A noteworthy mention is that our neural network was supervised training using sweat samples with known ion concentrations before the primary evaluation. In instances where the chip's output was inaccurate, adjustments were made, leading to modifications in the neural network's node weights. A groundbreaking feature in our approach is the on-chip training capability, which obviates the need for external software, marking a significant stride in the field.

\subsubsection{SNN Neurons hardware implementation}
\leavevmode\\
In \autoref{spik_neuron_fig}, the architecture of neural processing components in SNN hardware is depicted. Depending on the neuron models and encoding strategies used in SNNs, the design of these neural components can vary. As an example, when utilizing the LIF model combined with rate coding techniques, the neural components might include the subsequent submodules: 1)~\emph{Multiplier}, 2)~\emph{Accumulator}, 3)~\emph{Thresholds}, and 4)~\emph{Spiking Encoder}:

\begin{figure}[h]
    \centerline{\includegraphics[width=\columnwidth]{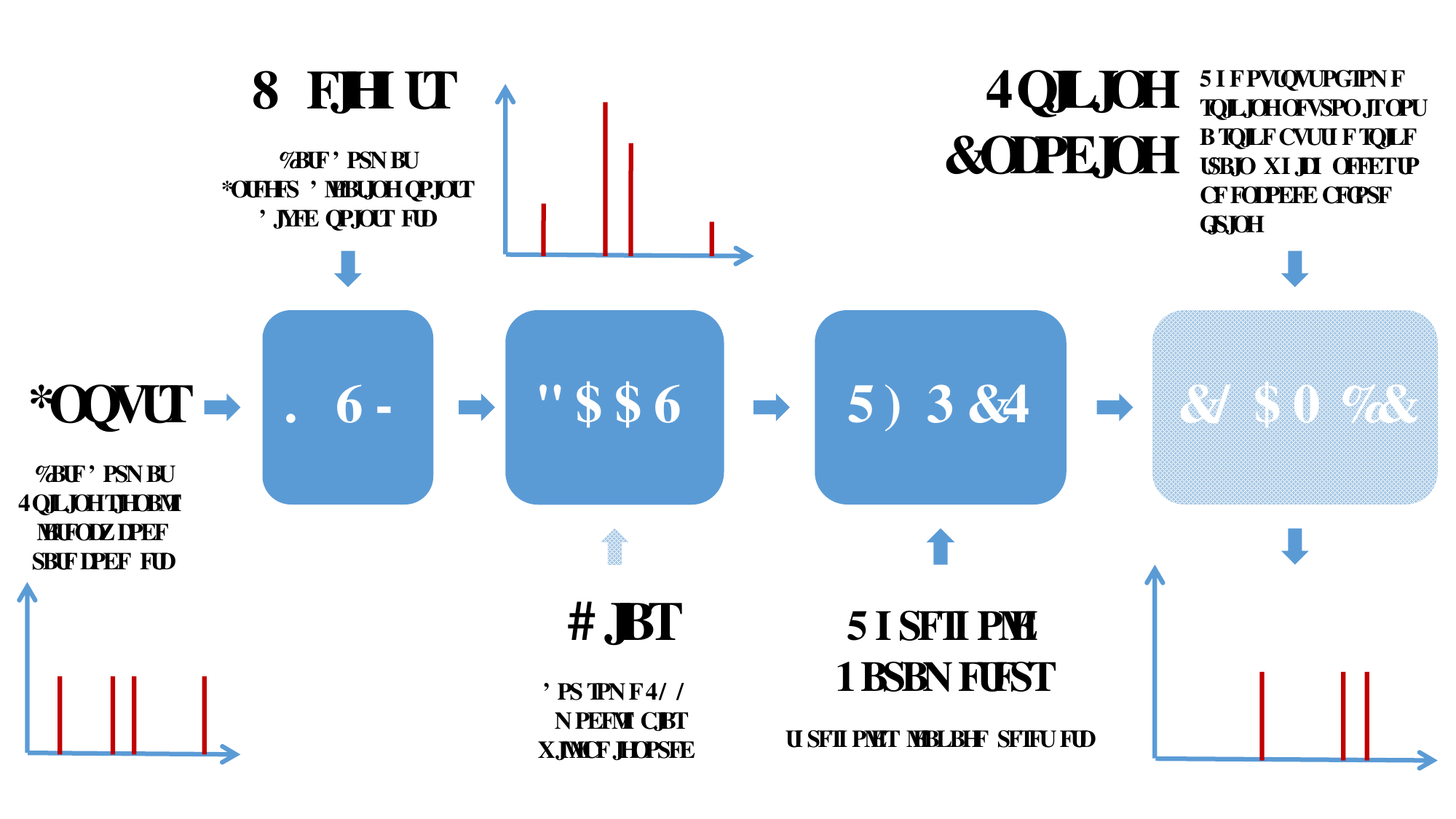}}
    \caption{Architecture of Spike-Based Neural Systems~\cite{isik2023survey}.}
    \label{spik_neuron_fig}
\end{figure}

\begin{itemize}
    \item \emph{\textbf{Multiplier}} and~\emph{\textbf{Accumulator}} compute aggregate of weighted spike values using varied data representations. Yet, retaining neural network weights demands significant on-board memory capacity within FPGA. Given hardware constraints, besides the usual floating-point and fixed-point representations, there's a growing interest in exploring reduced-precision data types in contemporary studies, including i)~\emph{Quantization}, ii)~\emph{Binarization}, iii)~\emph{Approximation Computation}, and iv)~\emph{Posit-floating Computation}:
    
    The authors of~\emph{Q-SpiNN}~\cite{putra2021q} and work~\cite{quan2020snn} implemented a~\emph{Quantized Spiking Neural Network} (QSNN) framework to reduce the memory consumption of network weights based on the low-precision integer scheme. Moreover, work~\emph{BS4NN}~\cite{kheradpisheh2021bs4nn} and work~\cite{wei2021binarized} further reduced the weight precision to the~\emph{Binarized Spiking Neural Network} (BSNN). Work~\emph{AxSNN}~\cite{sen2017approximate} explored the application of approximation computation on SNN. The authors of~\cite{silvaposit} researched the possibility of applying~\emph{Posit}-floating computation~\cite{posit1, posit2} on FPGA-based SNN accelerators.
    
    \item \emph{\textbf{Thresholds and Spiking Encoding}}: As depicted in~\autoref{spik_neuron_fig}, the criteria include factors like decay rate, recovery duration, activation limits, and so on. For certain encoding methods, like~\emph{Phase Coding}, an encoding component becomes essential.
\end{itemize}

\subsubsection{Coding Formats of Spiking Signals}
\leavevmode\\
In contemporary studies on artificial intelligence, spiking neural networks have emerged as a leading subject, aiming to emulate the operational mechanisms of biological neural systems. As a result, it's essential to transform input data into a format of spiking signals. For example, \autoref{spiking_signal_fig} displays five of the most commonly used spiking encoding methods in recent studies\cite{guo2021neural, von2017artificial, el2023efficiency}:

\begin{figure}[h]
    \centering\includegraphics[width=\columnwidth]{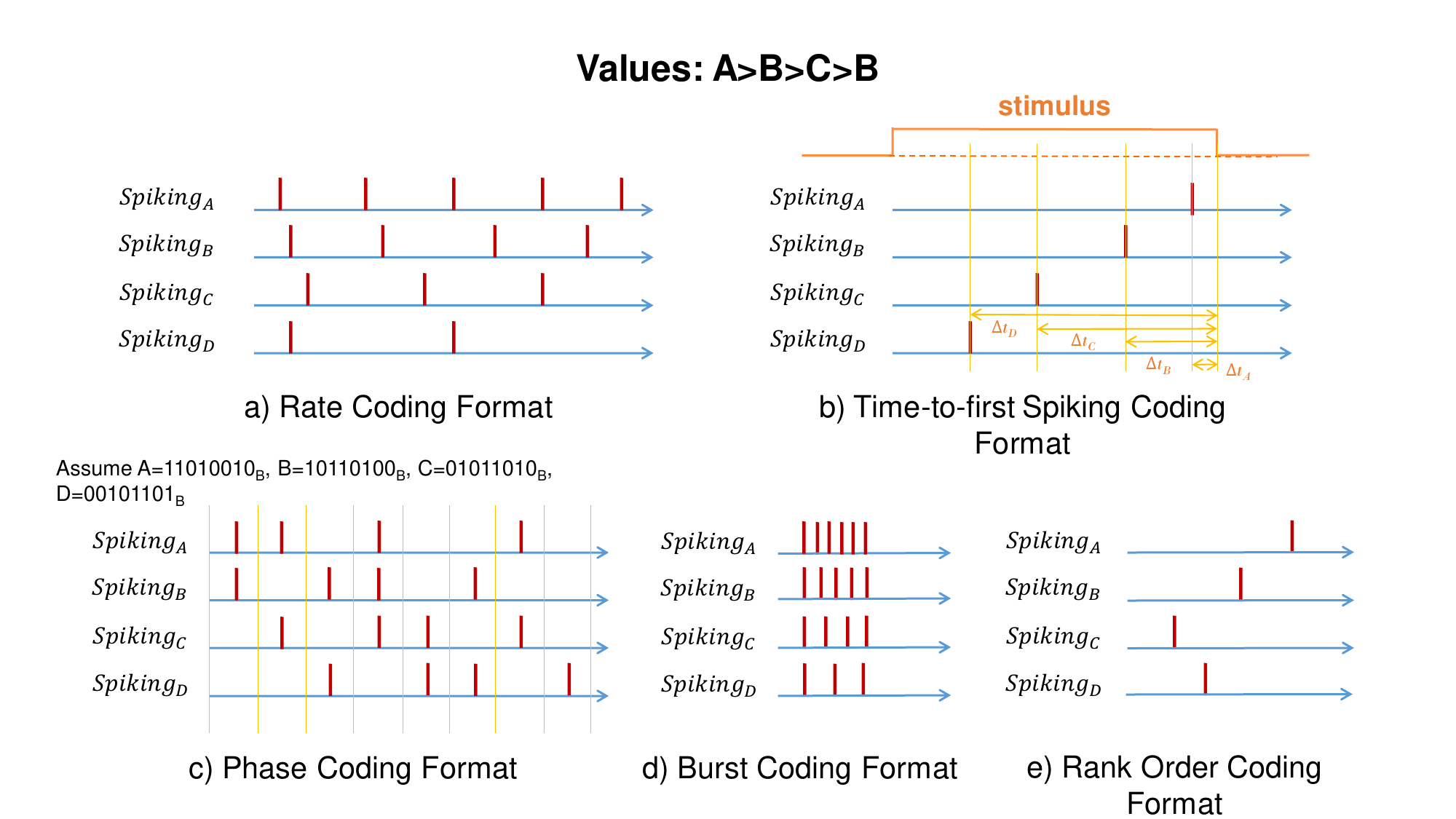}
    \centering\caption{Five popular spiking coding formats~\cite{isik2023survey} }
    \label{spiking_signal_fig}
\end{figure}

\begin{itemize}
    \item \emph{\textbf{Rate Coding Format: }} Rate coding is a predominant method in associated studies. This approach transforms the input data into a Poisson spike train that reflects the respective firing rate. As illustrated in~\autoref{spiking_signal_fig}a, a more substantial input is depicted as a spike train with an increased firing frequency.
    
    \item \emph{\textbf{Time-to-first Spiking Coding Format: }}~\emph{Time-to-first Spiking Coding} employs a temporal encoding technique. Rather than utilizing firing rate to depict varying input values, this method uses the firing delay within a stimulus period to represent data with a spike. As depicted in~\autoref{spiking_signal_fig}b, spikes corresponding to greater inputs occur sooner. Unlike the rate coding method, time-to-first coding needs just one spike, with data details encoded in the time delay.
    
    \item \emph{\textbf{Phase Coding Format: }} The two mentioned encoding techniques require a transformation between data values and firing rate or delay. The choice of conversion approach can affect the resulting spike trains. As illustrated in~\autoref{spiking_signal_fig}c, the study by~\cite{kim2018deep} investigated an alternative spiking encoding approach that directly translates the binary representation of an input value into a spike train. In this context, the weight of the spike signifies the importance of each input bit.
    
    \item \emph{\textbf{Burst Coding Format: }} To minimize the transmission time of spiking signals, the studies by~\cite{izhikevich2003bursts} and~\cite{eyherabide2009bursts} investigated an encoding technique involving burst transmission, as depicted in~\autoref{spiking_signal_fig}d. This approach represents the input value through the count of spikes and the interval between spikes (ISI) within the burst. A denser clustering of spikes in a burst indicates a higher input value.
    
    \item \emph{\textbf{Rank Order Coding Format: }}~\autoref{spiking_signal_fig}d shows the scheme of~\emph{Rank Order Coding}~\cite{thorpe1998rank}. Rather than using the spike latency in the stimulus for encoding as seen in the ~\emph{Time-to-first Spiking Coding} approach, this method captures input data based on the sequence in which spikes are fired, without considering the time delay between them. Consequently, for $N$ pre-neurons, the outputs can represent up to $2^N$ distinct data values.
\end{itemize}

\section{Transformers in Audio Processing}

\subsection{General Description of Transformers}
Transformers, introduced in \cite{vaswani2017attention}, have revolutionized the field of deep learning, particularly in natural language processing (NLP) tasks. The core strength of the Transformer architecture lies in its self-attention mechanism, which allows the model to weigh input data differently, focusing more on parts that are contextually relevant. This mechanism enables the model to capture long-range dependencies and relationships in the data, making it particularly powerful for sequence-to-sequence tasks.

\subsection{Transformers in Audio Processing}
While Transformers were initially designed for NLP tasks, their potential has been recognized in various other domains, including audio processing. In audio tasks, Transformers can capture temporal dependencies in audio sequences, making them suitable for tasks like speech recognition, audio synthesis, and sound classification. The self-attention mechanism allows the model to focus on specific parts of an audio clip, such as distinguishing speech from background noise or identifying specific instruments in a musical piece. Recent advancements have seen the adaptation of Transformer-based models, like BERT and GPT, for audio tasks, leading to state-of-the-art performance in several benchmarks.

\subsection{Hardware Implementation of Transformers}
The computational demands of Transformer models, especially when scaled, are significant. This has led to interest in hardware implementations that can accelerate their performance while reducing power consumption. Hardware platforms parallelize the matrix operations central to Transformer models, offering faster inference times. Moreover, with the advent of edge computing, there's a growing interest in deploying compact Transformer models on edge devices, necessitating efficient hardware implementations. Recent research has also explored quantization and pruning techniques to reduce the model size without significantly compromising performance, making them more amenable to hardware deployment.

\subsection{HPCNeuroNet}

\begin{figure*}[h]
    \centering
    \includegraphics[width = 0.7\textwidth]{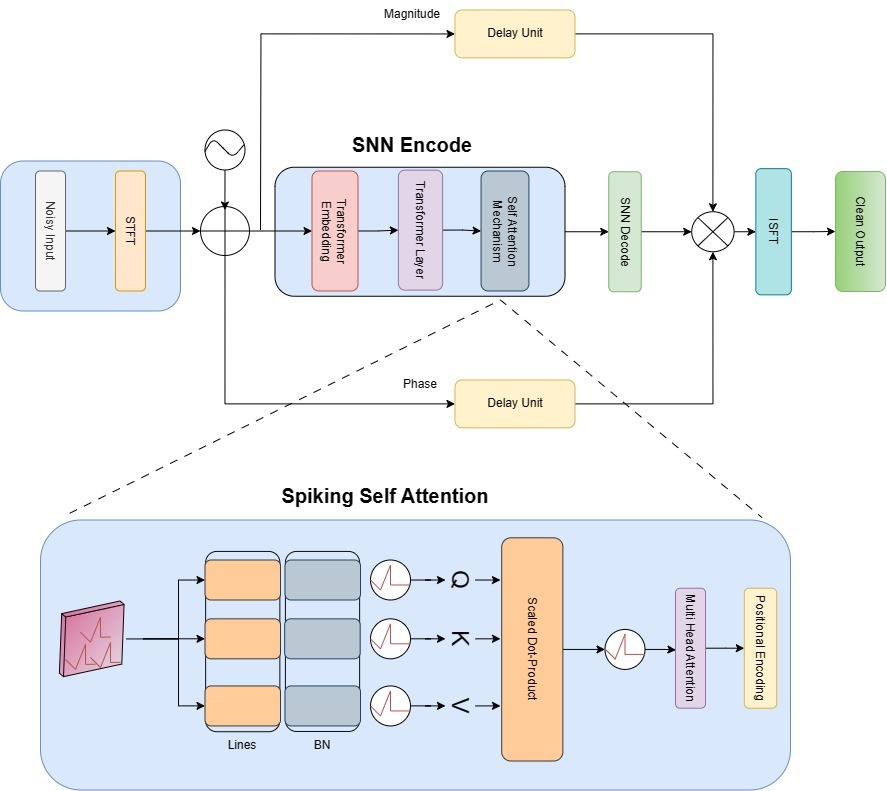}
    \caption{HPCNeuroNet Model Architecture. The left showcases the STFT transforming raw audio into a time-frequency representation. The middle highlights the Transformer Embedding layer and the core HPCNeuroNet Transformer Layers with mechanisms like Self-Attention and Scaled Dot-Product Attention. The Spiking Self Attention, SNN Encode, and SNN Decode layers introduce and process SNN dynamics. The final Clean Output is refined using Magnitude, Phase, and ISTFT, with the Delay Unit ensuring temporal synchronization.}
    \label{Spiking Transformer Model Architecture}
\end{figure*}

HPCNeuroNet is a cutting-edge architecture designed to seamlessly integrate the computational prowess of HPC with the temporal dynamics of SNN and the attention mechanisms of Transformers, specifically tailored for audio processing which are shown in~\autoref{Spiking Transformer Model Architecture}. The audio data, in its raw waveform, is first transformed using the \textit{Short-Time Fourier Transform (STFT)} to convert it into a time-frequency representation. This representation captures both the temporal and frequency characteristics of the audio, making it suitable for further processing. The transformed data, termed the \textit{Noisy Input}, represents the audio with potential noise or unwanted signals. This input is then passed to the \textit{Transformer Embedding} layer, which converts the time-frequency representation into dense vector embeddings. These embeddings encapsulate the salient features of the audio data and are primed for processing by the subsequent Transformer layers. The core computational layers of the architecture are the \textit{HPCNeuroNet Transformer Layers}. These layers process the embeddings, discerning patterns, and relationships intrinsic to the audio data. Within these layers, the \textit{Self-Attention Mechanism} plays a pivotal role. It allows the model to focus on different parts of the audio data based on their relevance. This mechanism involves transformations into Query (Q), Key (K), and Value (V) matrices, followed by \textit{Scaled Dot-Product Attention}, \textit{Multi-Head Attention}, and \textit{Positional Encoding}. After processing through the Transformer layers, the data is passed to the \textit{Spiking Self Attention} mechanism, which introduces the temporal dynamics of SNN into the attention mechanism. The output from this layer is then encoded into spike trains using the \textit{SNN Encode} layer, leveraging mechanisms such as Rate Coding and Time-to-First-Spike Coding. The spike-encoded data is then decoded using the \textit{SNN Decode} layer, transforming the spike trains back into a time-frequency representation. This representation undergoes an \textit{Inverse Short-Time Fourier Transform (ISTFT)}, converting it back to the time-domain. The \textit{Clean Output} represents the denoised or processed audio data, with unwanted signals or noise mitigated. This output is then further refined using \textit{Magnitude} and \textit{Phase} information, ensuring the audio's fidelity and clarity. The \textit{Delay Unit} introduces temporal dynamics into the network, ensuring that the network's response is synchronized with the temporal characteristics of the audio data. The entire process culminates in the \textit{End}, marking the completion of the audio data's journey through the HPCNeuroNet architecture and producing a processed audio output that is enhanced, denoised, and ready for playback or further analysis.
    
Table \ref{tab:HPCNeuroNet characteristics} delineates the salient characteristics of the HPCNeuroNet system. Each row of the table represents a specific configuration or instantiation of the system, and the columns provide detailed specifications for various parameters.

\begin{itemize}
    \item \textbf{Number of Bands}: This column specifies the number of frequency bands into which the input audio signal is decomposed. In the given configuration, the audio signal is divided into 256 distinct frequency bands, allowing for a granular representation of the signal in the frequency domain.
    
    \item \textbf{Frequency Range}: This column indicates the range of frequencies that the system can process. The provided configuration is capable of processing signals ranging from 0 Hz to 8 kHz, encompassing most of the human audible range and ensuring that a wide variety of audio signals can be effectively handled.
    
    \item \textbf{Dynamic Range}: The dynamic range, measured in decibels (dB), represents the ratio between the largest and smallest values that the system can process. A dynamic range of 96 dB signifies that the system can handle a vast range of amplitude variations, from very soft to very loud sounds, without distortion or loss of fidelity.
    
    \item \textbf{Max. Event Rate}: This metric, expressed in Mevents/s (Million events per second), provides an estimate of the system's processing capability in terms of the maximum number of events it can handle per second. An event rate of 8.76 Mevents/s indicates a robust processing capability, ensuring real-time or near-real-time processing of input signals.
    
    \item \textbf{Clock Frequency}: The clock frequency, given in MHz, denotes the operating frequency of the system's core components. A clock frequency of 250 MHz ensures rapid processing and timely response to input signals, facilitating real-time applications and efficient system performance.
\end{itemize}

\begin{table*}[h]
    \centering
        \caption{HPCNeuroNet characteristics}
    \begin{tabular}{|l|l|l|l|l|}
        \hline
        \textbf{Number bands} & \textbf{Frequency range} & \textbf{Dynamic range} & \textbf{Max. Event rate} & \textbf{Clock frequency} \\
        \hline
         256 & 0 Hz–8 kHz & 96 dB & 8.76 Mevents/s & 250 MHz \\
        \hline
    \end{tabular}

    \label{tab:HPCNeuroNet characteristics}
\end{table*}

\subsubsection{CPU/GPU Implementation}
\leavevmode\\
We utilized Python to execute implementations on the CPU and GPU. The study leveraged the computational prowess of NVIDIA's GeForce RTX 3060 GPU and Intel's Core i9 12900H CPU, both of which are optimized for different tasks, ensuring an efficient execution of our implementations.

\subsubsection{FPGA Implementation}
\leavevmode\\
\begin{figure}[h]
    \centering
    \includegraphics[width = 0.5\textwidth]{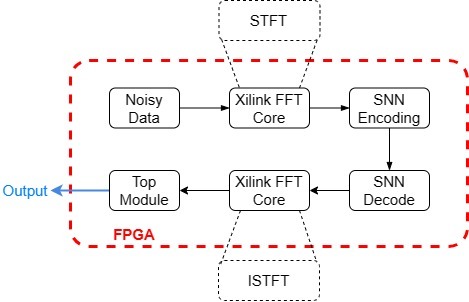}
    \caption{Hardware architecture of HPCNeuroNet.}
    \label{Hardware}
\end{figure}
The architecture design implements the various processes in parallel, thus consuming less processing time. Xilinx Intellectual Property (IP) cores and custom modules are used in the implementation which is illustrated in~\autoref{Hardware}. Modules are defined using VHDL based on the AXI Interface. The modules in the design use \(16\)-bit data configuration for the real and imaginary parts. Xilinx FFT core is used to improve the efficiency and speed of the STFT and ISTFT blocks which have a maximum instantaneous frequency of \(250\) MHz and can operate up to \(50\) frames/sec. We used the implementation according to dataset is 48000  1x 1D vectors.

\subsection*{\textbf{STFT:}}
STFT is a pivotal component in the HPCNeuroNet architecture, responsible for converting the raw audio waveform into a time-frequency representation. Given the dataset's structure, with each audio sample being a 1D vector of length 48000, the FFT core was configured accordingly. The STFT operation uses a window length of 512 with a hop length of 128, leading to 8 ms per time-step, given the signal's 16 kHz sampling rate. This configuration ensures that both the temporal and frequency characteristics of the audio are captured efficiently.

\subsection*{\textbf{SNN Encode:}}
The encoding process converts the time-frequency representation into spike trains, optimized for speed and efficiency. The \(16\)-bit data configuration ensures that the spike trains capture the nuances of the audio data while being suitable for real-time processing. The encoder's design is influenced by the neuromorphic denoising network, ensuring efficient operation and synergy with the subsequent processing stages.

\subsection*{\textbf{SNN Decode:}}
The \(16\)-bit data configuration is maintained, ensuring that the decoded time-frequency representation retains the audio data's original characteristics, albeit processed. The decoder's design is tailored to operate in conjunction with the encoder, ensuring seamless integration and efficient data flow within the HPCNeuroNet architecture.

\subsection*{\textbf{ISTFT:}}
ISTFT is responsible for converting the time-frequency representation back into the time-domain audio waveform. Similar to the STFT, the core operates with a \(16\)-bit data configuration for both the real and imaginary parts, ensuring precision in the reconstructed audio. The ISTFT operation mirrors the STFT's parameters, using the same window length and hop length. The resulting output is the processed audio waveform, ready for playback or further analysis.

\vspace{5pt}
The concept of design-space exploration (DSE) has gained traction in neuromorphic hardware research, offering insights into optimizing audio processing tasks. The judicious selection of core capacity is paramount to achieving optimal performance and energy efficiency metrics.

\subsubsection{Hardware Accelerator MAC Components}
\leavevmode\\
The number of Multiply-Accumulate (MAC) operations in the HPCNeuroNet architecture can be quantified for its primary components:

\begin{enumerate}
    \item \textbf{Transformer Embedding}: The embedding layer involves matrix multiplications to convert input data into dense vector embeddings, where \(D\) represents the number of input features or dimensions in the input data and \(S\) denotes the embedding dimension, which is the number of dimensions in the dense vector.
    \begin{equation}
        \text{MAC}_{\text{Embedding}} = D \times S
    \end{equation}

    \item \textbf{HPCNeuroNet Transformer Layers}: The transformer's self-attention mechanism involves matrix multiplications for the Query (Q), Key (K), and Value (V) transformations, as well as the Scaled Dot-Product Attention, where \(T\) represents the number of attention heads in the Multi-Head Attention mechanism of the transformer layers.
    \begin{equation}
        \text{MAC}_{\text{Transformer}} = 2 \times D^2 \times S + D \times S^2 \times T
    \end{equation}

    \item \textbf{Spiking Self Attention and SNN Encode/Decode}: The MAC operations for the SNN layers, where \(N\) signifies the number of neurons/processing units in SNN and \(t_{\text{sim}}\) represents simulation time.
    \begin{equation}
        \text{MAC}_{\text{SNN}} = N \times t_{\text{sim}}
    \end{equation}
\end{enumerate}

The total MAC operations for the HPCNeuroNet architecture is the sum of the MAC operations of its components:
\begin{equation}
    \text{MAC}_{\text{Total}} = \text{MAC}_{\text{Embedding}} + \text{MAC}_{\text{Transformer}} + \text{MAC}_{\text{SNN}}
\end{equation}

\subsubsection{Throughput Analysis}
\leavevmode\\
The MAC counts, as derived from neural network libraries \cite{WinNT6, WinNT7}, and the latency or simulation time, primarily influenced by the hardware components, can be used to compute the throughput of the system:

\begin{equation}
\text{Simulation Time} = \frac{\text{Total Inference Time}}{\text{Total number of inference samples}}
\label{eq:sim}
\end{equation}

This relationship can be represented as:

\begin{equation}
    \text{Throughput} = \frac{\# \text{MACs}}{\text{Latency (Simulation Time)}}
    \label{eq:throughput}
\end{equation}

\subsubsection{Hit rate of sound recognition system}
\leavevmode\\
The efficacy of the N-DNS Challenge is gauged by the clarity of the produced audio. We utilize the Scale-Invariant Source-to-Noise Ratio (SI-SNR) to evaluate this clarity. This metric is widely recognized in the domain of audio processing, as highlighted in references such as \cite{bahmaninezhad2019comprehensive, le2019sdr}. The SI-SNR evaluates the prominence of human speech over background noise in the system's output, drawing parallels with the conventional Source-to-Noise Ratio (SNR) \cite{le2019sdr}. A distinguishing attribute of SI-SNR is its immunity to changes in amplitude. In essence, this ensures that the metric remains unbiased, regardless of variations in output loudness. Given an input waveform, denoted by a zero-mean vector $s$, and its corresponding system output $\hat{s}$ and \( e_{\text{noise}} \) represents the noise or error in the signal, the SI-SNR is expressed as:

\begin{equation}
\text{SI-SNR} := 10 \log_{10} \left( \frac{||s_{\text{target}}||^2}{||e_{\text{noise}}||^2} \right),
\end{equation}

with

\begin{equation}
s_{\text{target}} := \frac{\langle \hat{s}, s \rangle s}{||s||^2} \quad \text{and} \quad e_{\text{noise}} := \hat{s} - s_{\text{target}}.
\end{equation}

Our preference for SI-SNR stems from its straightforwardness and broad applicability, contrasting with more complex metrics like the speech-to-text accuracy in the Microsoft DNS Challenge \cite{dubey2022icassp}. The N-DNS challenge primarily seeks to promote advancements in neuromorphic techniques. We view the audio denoising task as representative of typical audio processing tasks. While some commercial ventures might not emphasize perceptual quality for human listeners, SI-SNR offers a practical metric that can also serve as a loss function in machine learning approaches.

\section{Results}

\begin{table*}[h]
\small
\centering
\caption{Evaluation results on CPU, GPU, and our processor.}
\label{table:table_2}
\begin{tabular}{|c|c|c|c|c|c|c|c|}
\hline
&\begin{tabular}[c]{@{}c@{}}Technology\\ {[}nm{]}\end{tabular} & \begin{tabular}[c]{@{}c@{}}Frequency\\ {[}MHz{]}\end{tabular}& \begin{tabular}[c]{@{}c@{}}\# of MAC\\ {[}MOP{]}\end{tabular}&\begin{tabular}[c]{@{}c@{}}Latency\\ {[}ms{]}\end{tabular} & \begin{tabular}[c]{@{}c@{}}Throughput\\ {[}GOP/s{]}\end{tabular}  &\begin{tabular}[c]{@{}c@{}} Power {[}Watt{]}\end{tabular}& \begin{tabular}[c]{@{}c@{}}Power Efficiency\\ {[}GOP/s/W{]}\end{tabular} \\ \hline
Intel i9 12900H (CPU) & 10 & 3700& 960& 110.62 & 8.67&28 &0.30 \\ \hline
NVIDIA RTX 3060 (GPU) & 8 & 1320& 960& 3.15 & 304.76&80 &3.80\\ \hline
Xilinx VU37P (FPGA) & 16 & 100& 960& 13.5 & 71.11&3.55 &19.75\\ \hline

\end{tabular}
\label{tab:comparison}
\end{table*}

Table \ref{tab:comparison} covers a number of key metrics, including manufacturing technology, operating frequency, power consumption, and other parameters such as latency, throughput, and power efficiency. This is evident when examining the latency metric for the Xilinx FPGA, which stands at a mere 13.5 ms. In terms of power consumption, the FPGA demonstrates remarkable energy efficiency with a power requirement of only 3.55 Watts, significantly less than both the CPU and GPU. The table further highlights the performance-per-watt of the FPGA with a throughput of 71.11 GOP/s and power efficiency of 19.75 GOP/s/W, underlining the suitability of FPGA devices for tasks where power efficiency is critical. This comparison reveals the distinctive characteristics and advantages of each technology, and their appropriateness would largely depend on the specifics of the application at hand.

%Table 1
\begin{figure}[h]
        \centerline{\includegraphics[width =\columnwidth]{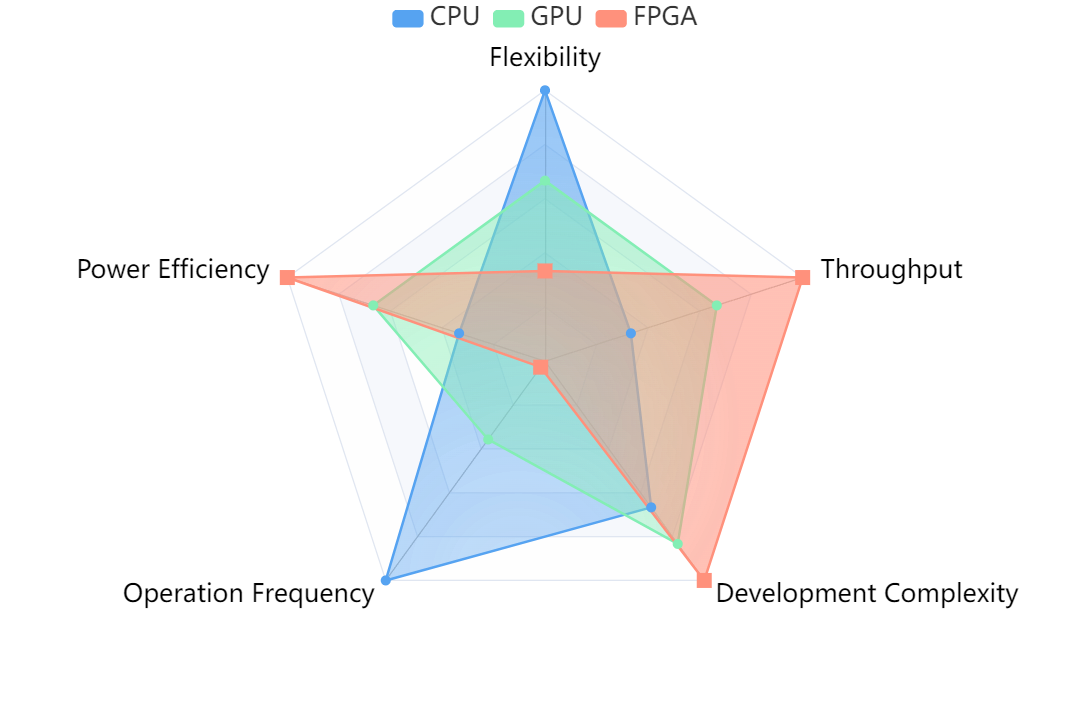}}
        \caption{HPCNeuroNet on Different Heterogeneous Devices}
        \label{Chart1}
    \end{figure}

~\autoref{Chart1} shows that CPUs, with their general-purpose design, offer unparalleled flexibility, making them suitable for a diverse range of tasks. However, this versatility often comes at the cost of power efficiency and throughput, especially when compared to specialized hardware like GPUs and FPGAs. GPUs, originally designed for graphics rendering, have been repurposed for parallel computing tasks due to their many-core architecture, granting them higher throughput for specific parallelizable tasks. Yet, they still lag behind FPGAs in terms of power efficiency. FPGAs, with their reconfigurable nature, allow for custom-tailored hardware designs that can achieve superior power efficiency and throughput for specific applications. However, this customization introduces a higher development complexity, requiring expertise in digital design and domain-specific knowledge. In essence, while CPUs provide broad applicability, GPUs and FPGAs offer specialized performance advantages, each with its trade-offs in terms of efficiency, throughput, and development complexity.

\begin{table*}
\centering
\tiny
\caption{Comparison of Results on Audio Processing Tasks}
\rotatebox{90}{% <-- This rotates the table counterclockwise

\footnotesize
\begin{tabularx}{1.2\textwidth}{|X|X|X|X|X|X|X|X|X|X|X|}
\toprule
\textbf{Work} & \textbf{NN} & \textbf{HW} & \textbf{Platform} & \textbf{Freq.} & \textbf{MAC Number (OP)} & \textbf{SNR Rate (db)} & \textbf{Power (W)} & \textbf{Latency (ms)} & \textbf{Throughput (GOP/s)}& \textbf{Scaled Efficiency (GOP/s/W)} \\
\midrule
\cite{ko2017precision}  & CNN-LSTM & CPU & Intel i7-9750H & 2.60GHz & 12.5M & N/A & 25 & 20 & 6.25 &0.25\\
\hline
\cite{ko2017precision}  & CNN-LSTM & GPU & NVIDIA RTX 2080 & 1.6GHz & 135M & N/A & 250 & 5 & 27 &0.108\\
\hline
\cite{yoshimura2020end} &Transformer & GPU & NVIDIA RTX 2080 & 1.6GHz & 13.5G & N/A & 250 & 20 & 675 &2.7\\
\hline
\cite{wang2018c}  & LSTM & FPGA & Xilinx Virtex-7 & 200MHz & N/A & N/A & 22 & 16.7us & 13.18 &0.599\\
\hline
\cite{gao2019real}  & DRNN & FPGA & Xilinx Zynq-7100 & 100MHz & N/A & [0, 5, 10] & 2.78 & 7.90 & 7.22 &2.597\\

\hline

\cite{diamos2016persistent}  & LSTM & GPU & NVIDIA TitanX & 1.15 GHz & 4290M & 25.0 & 150 & 10 & 429 &2.86\\
\hline
\cite{deng2021reconstruction}  & SNN & FPGA & Intel Cyclone-4 & 65.03 MHz & N/A & 25.0 & 1.684 & N/A & 0.71 &0.42\\
\hline
\cite{deng2023auditory} & SNN & FPGA & Intel Cyclone-4 & 107.28 MHz & N/A & 25.0 & 5.148 & 0.27 & 53.6 & 10.42\\
\hline

\cite{timcheck2023intel}  & Sigma-Delta NN & ASIC  & Intel Loihi 2 & 1 kHz & N/A & 12.71 & N/A & 0.36 & N/A &N/A\\
\hline
\textbf{Ours}  & \textbf{Hybrid (SNN\& Transformer)} & \textbf{FPGA} & \textbf{Xilinx VU37P} & \textbf{100MHz} & \textbf{960M} & 12.14 &\textbf{3.55} & \textbf{13.5} & \textbf{70.11} &\textbf{19.75}\\

\bottomrule
\end{tabularx}
}

\label{table:table_3}
\end{table*}

As shown in Table \ref{table:table_3}, we compare our Hybrid (SNN\&Transformer) implementation with other audio processing models presented in the literature. The evaluation of our design is done on the Xilinx VU37P FPGA platform operating at 100MHz. Our experimental results show that the proposed accelerator achieves a throughput of 70.11 GOP/s with a 3.55 W on-chip power consumption. The SNR for our model is 12.14 dB. When compared with off-the-shelf devices and recent state-of-the-art implementations, our proposed accelerator demonstrates significant advantages in terms of energy efficiency, achieving a scaled efficiency of 19.75 GOP/s/W, and offers design flexibility.

\section{Conclusion}
The integration of neuromorphic systems with HPC has ushered in a new era of audio data processing. This confluence has not only streamlined processing but also enhanced efficiency and speed, making significant strides in real-time audio applications. Our research, anchored by the results presented, emphasizes the pivotal role of HPC in augmenting the capabilities of a Hybrid (SNN \& Transformer) model for neuromorphic audio processing. The salient contributions of HPC to neuromorphic audio processing, as evidenced by our study, are manifold. The remarkable throughput and energy efficiency metrics achieved in our experiments underscore the transformative potential of HPC when combined with neuromorphic principles. The fusion of SNNs with transformers, in particular, heralds a novel approach to neuromorphic audio processing. This synergy capitalizes on the strengths of both paradigms, paving the way for unprecedented energy efficiency, real-time processing capabilities, and a more nuanced representation of audio data. The transformative potential of neuromorphic hardware in the domain of audio processing, when bolstered by the computational prowess of HPC, is undeniable. As we stand at this juncture, the horizon is replete with opportunities, and the future beckons with the promise of innovations that will redefine the landscape of neuromorphic audio processing.

\section{Future Work}
The integration of SNNs with transformers introduces complexities, notably in the realms of training dynamics and system scalability. However, the anticipated benefits underscore its potential as a promising avenue in neuromorphic audio processing research. Subsequent investigations will aim to assess its practicality in real-world audio processing, specifically within innovative TEMSA Skoda vehicles that are equipped with neuromorphic audio systems.

\section*{Acknowledgments}

The authors would like to thank to Mert Hidayetoglu and Jonathan Naoukin for the feedback, which made the attainment of the results reported in this paper possible.

\bibliographystyle{IEEEtran}
\bibliography{sample-base}

\end{document}